# GridBank: A Grid Accounting Services Architecture (GASA) for Distributed Systems Sharing and Integration

**Alexander Barmouta**

Department of Computer Science and Software Engineering
University of Western Australia
Nedlands, Western Australia, 6009

`barmouta@csse.uwa.edu.au`

**Rajkumar Buyya**

Grid Computing and Distributed Systems (GRIDS) Lab
School of Computer Science and Software Engineering
University of Melbourne
Melbourne, Victoria, Australia

`raj@cs.mu.oz.au`

**Abstract**

Computational Grids are emerging as new infrastructure for Internet-based parallel and distributed computing. They enable the sharing, exchange, discovery, and aggregation of resources distributed across multiple administrative domains, organizations and enterprises. To accomplish this, Grids need infrastructure that supports various services: security, uniform access, resource management, scheduling, application composition, computational economy, and accountability. Many Grid projects have developed technologies that provide many of these services with an exception of accountability. To overcome this limitation, we propose a new infrastructure called Grid Bank that provides services for accounting. This paper presents requirements of Grid accountability and different models within which it can operate and proposes Grid Bank Services Architecture that meets them. The paper highlights implementation issues with detailed discussion on format for various records/database that the GridBank need to maintain. It also presents protocols for interaction between GridBank and various components within Grid computing environments.

## 1   Introduction

As the trend towards Internet-based distributed computing continues, large scale computationally intensive applications are executed on remote machines that are offered to provide computational services during periods when these computers are idle. Hosts connected by Internet with middleware supporting remote submission and execution of applications constitute what is called *the computational Grid* [2,9,10,12]. The Grid couples a variety of heterogeneous computational resources, storage systems, databases and other special-purpose computing devices and presents them as a unified integrated resource. In the global Grid environment users submit their applications to Grid Resource Broker, which discovers resources, negotiates for service costs, performs resource selection, schedules tasks to resources and monitors task executions. Resource providers advertise their services with the discovery service and run Grid Trade Service used by Grid Resource Broker to negotiate service cost. Such setup allows open market trade of computational services to take place on the Grid. Resources are offered at difference prices, and those prices are negotiated using one of several economic models from the real world [2,4].

It was observed that the utility delivered by resources is enhanced when resource allocation is performed based on users quality-of-service (QoS) requirements/constraints (e.g., deadline and budget) [2]. In a global computing environment all users would prefer to use powerful resources, which would cause some resources to be oversubscribed and others undersubscribed. This is where computational economy and suitable service pricing strategies come into play. Resource owners are permitted to solicit an open market price in a way that achieves maximum profit and resource consumers are allowed to choose resources that meet their QoS requirements. That is, when there is less demand for resources, the price is lowered; when there is high demand, the price is raised. This helps in regulating the supply-and-demand for access to Grid resources and services.

The Gridbus project [3] is developing technologies that enable service-oriented cluster and grid computing. It is driven by a distributed computational economy approach to the resource sharing, exchange, discovery, and aggregation. *GRid Architecture for Computational Economy* (GRACE) is a generic framework allowing creation of the World-Wide Grid marketplace [2,4]. This infrastructure includes components for resource selection and scheduling:

- Nimrod-G (Grid Resource Broker designed for parameterized applications)
- Grid Resource and Market Information Server
- Grid Open Trading protocols and APIs
- Grid Trade Manager (part of broker involved in establishing service price)

- Grid Trade Server (for resource owners)

The Gridbus project aims to realize the full potential of GRACE framework and demonstrate its capability at various levels: cluster, peer-to-peer (P2P) networks, and Grid. At cluster level, Libra technology has been developed for economy-driven cluster scheduling [3]. It is used within single administrative domain for distributing computational tasks among resources that belong to a cluster. At P2P network level, the CPM (compute-power-market) infrastructure is being developed through Jxta community. At Grid level, various tools are being developed to support QoS based schedule for both compute and data-intensive applications. "GridSim" is a Grid simulation toolkit for resource modeling and application scheduling, which can be used to simulate rather than build a computational Grid for testing purposes. In addition, to support accountability and sustained resource sharing across various organizations, Gridbus project and we are developing GridBank infrastructure, which is discussed in rest of this paper.

At present, access to resources is not economy-based. Resource owners and users co-operate as part of the same *Virtual Organization* (VO) [10,12], which is a collaboration of people agreed to share resources. In order to share computational services across multiple VOs (administrative domains) there is a need for accounting infrastructure that would allow unambiguous recording of user identities against resource usage. In addition, VOs can choose to introduce their own currency for resource trading. In the context of Gridbus project we call such system the *GridBank*.

GridBank is a secure Grid-wide accounting and (micro) payment handling system. It maintains users (both consumer and provider) accounts and resource usage records in the database. It supports protocols that enable its interaction with Grid Service Consumers (GSCs) resource broker and Grid Service Providers (GSPs). It has been primarily envisioned to provide services for enabling Grid computing economy; however, we envision its usage in E-commerce applications. The GB services can be used in both co-operative and competitive distributed computing environments.

GridBank can be thought of as just another resource on the Grid. In other words, it is just another Grid Service Provider. Clients use the same user proxy/component to access GridBank as they use to access other resources on the Grid. A user proxy is a certificate signed by the user, which is later used to repeatedly authenticate the user to resources [7,9,15]. This preserves Grid's single sign-in policy and avoids repeatedly entering user password. Using existing payment systems for the Grid would not satisfy this policy.

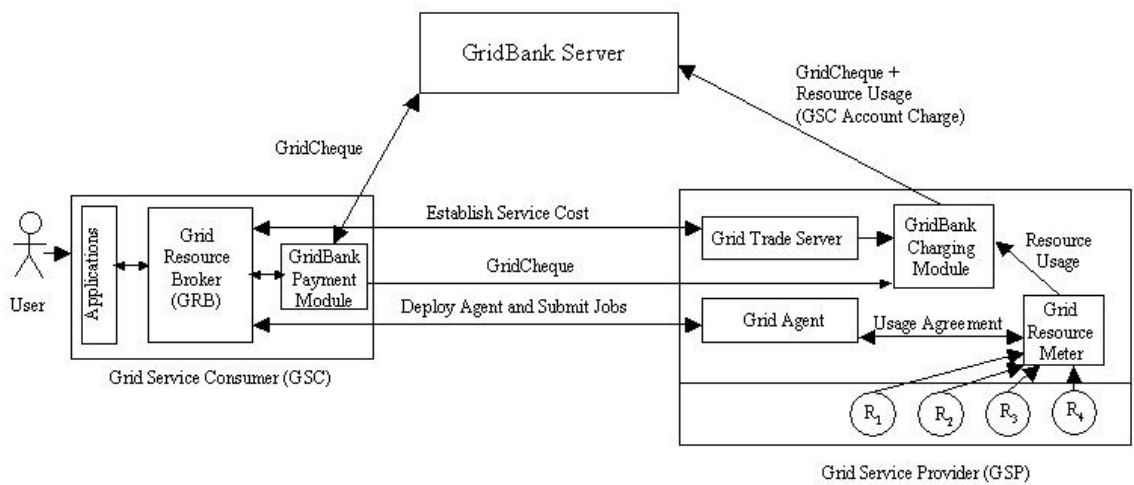

Figure 1: Interaction of GridBank with other Grid components.

## 2 Components Interaction

**Use Case: The interaction between GridBank and various components of Grid is shown in** Figure 1. GSPs and GSCs open account with GridBank. Then, the user submits application processing requirements along with QoS requirements (e.g., deadline and budget) to the Grid Resource Broker (GRB). The GRB interacts with GSP's Grid Trading Service (GTS) or Grid Market Directory (GMD) to establish the cost of services and then selects suitable GSP. It then submits user jobs to the GSP for processing along with details of its chargeable account ID in the GridBank or GridCheque purchased from the GridBank. The GSP provides the service by executing the user job and the GSP's Grid Resource Meter measures the resources consumed while processing the user job. The GSP's charging module contacts the GridBank with request to charge the user account and transfer the funds/credits to the GSP account. It also passes information related to the reason for charging (resource usage record).

### 2.1 GSP Components Interaction

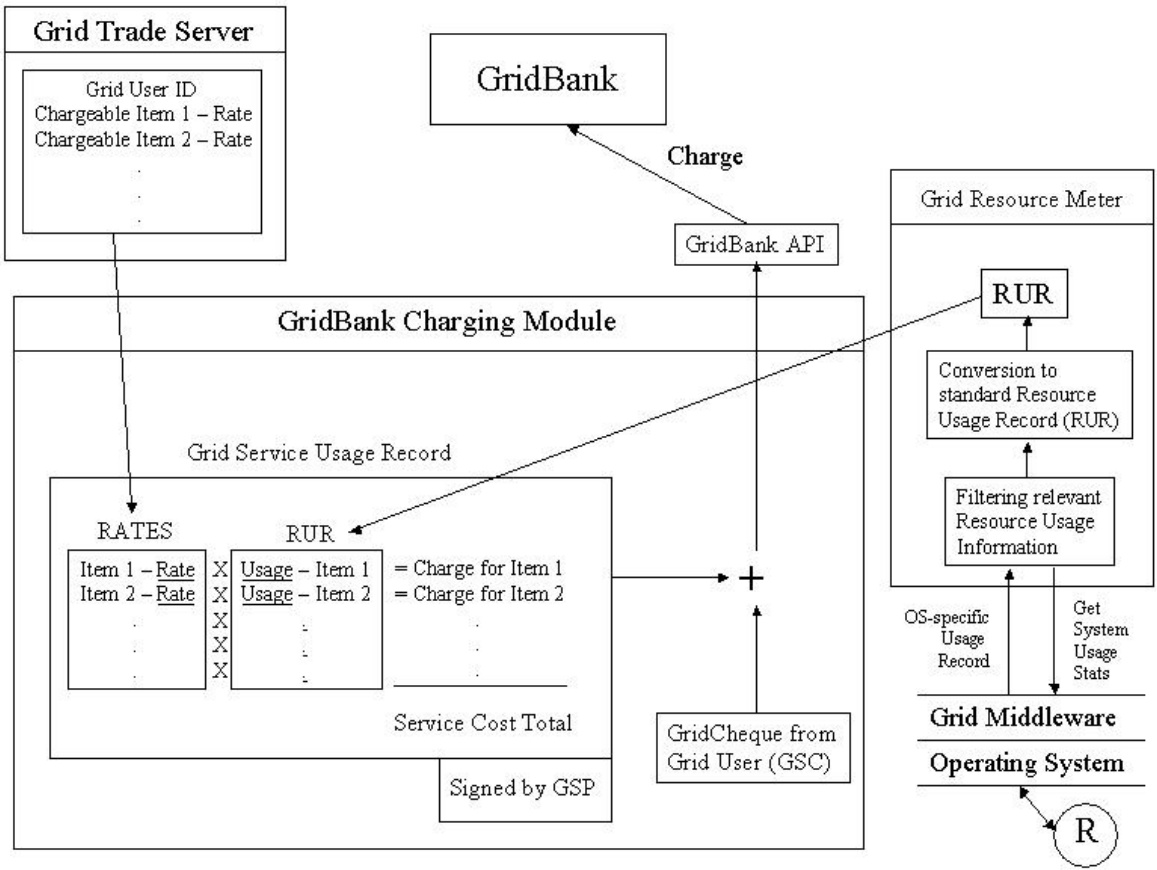

Figure 2: Interaction between GTS, GBCM, GRM and GridBank.

The *Grid Resource Meter* (GRM) module will interface with local resource allocation system (e.g., cluster scheduler) or user-level Grid agents (such as Nimrod-G agent) to extract resource usage information. The interfacing can be mediated through Grid middleware tools such as Globus [8], by extending it such that a native operating system usage function is called after a user application finished execution, and another function that is called by GRM to collect the data. Once GRM obtains the raw usage statistics (see Figure 2), it filters relevant fields in the record and passes them to the conversion unit, which generates a standard OS-independent *Resource Usage Record* (RUR). The RUR is a combined effort of the Grid community at the Global Grid Forum [13,20]. GridBank stores this record in its database, which provides evidence that a transaction took place. RUR is then forwarded to GridBank Charging Module (GBCM) for calculation of the total service cost. GBCM obtains service rates for the user from the Grid Trade Server (GTS). The user ID (Certificate Name) passed to GBCM is contained in the GridCheque (or any other payment instrument). The service rates record generated by the GTS and the Resource Usage Record must conform to each other. For every chargeable item in the rates record there must be a corresponding item in the RUR. Chargeable items to be considered are [2,4] :

- Processors: User CPU time

- Main Memory
- Secondary Storage
- I/O channels (such as networking)
- Software Libraries: System CPU time

The total charge is calculated by multiplying rate by usage for each item and then adding up individual charges. The rate for CPU time is G$ (Grid currency) per CPU hour and the usage is time. The rate for memory and storage is G$ per MB*hour and usage is MB*hour. I/O service rate is G$ per MB and usage is MB of total "traffic". These calculations along with the rates and RUR records are signed by GSP to provide non-repudiation of the transaction, and are submitted together with GridCheque (or other payment instrument) to GridBank Server for processing.

Besides the basic functionality, the GRM provides different levels of accounting information depending on the kind of payment protocol GridBank Charging Module is using. Different protocols might require different resource usage statistics. For example, in Figure 1, each individual resource ($R_1 – R_4$) used to provide computational service presents its usage record to Grid Resource Meter (GRM). GRM might choose to aggregate individual records into the standard RUR to reflect the charge for the combined GSP's service.

### 2.2 GSC Components Interaction

Grid Resource Broker (in our implementation Nimrod-G resource broker [2,5]) performs service cost (rates) negotiations with GSP's Grid Trade Server and deploys the Grid Agent responsible for setting up execution environment on GSP's machine and downloading the application and data from remote locations if they are not already on the machine [2]. However, before the broker can submit a job, a local account on the remote host must exist [15]. Section 2.3 addresses this issue.

GRB interacts with GridBank Payment Module to manage funds on user's behalf. The user can then set the budget to prevent overspending. GRB submits a job for execution on the resource in similar fashion as normally (using Globus's submit-job command [15]), but the call is made using GBPM's API such that it initially forwards payment details to GridBank Charging Module in order to authorize access to the service.

### 2.3 Access Scalability

Grid middleware technologies such as Globus assume that local access to resources is managed by the resource owner who has to create and manage local accounts for each user. Thousands (or even millions) of GSCs can be clients of GridBank and the requirement to have a local account at each resource is simply not realistic. GridBank Charging Module alleviates the problem. This can be achieved by enhancing GridBank to provide "arbitration" and "resource access authorization" services like a market-maker that brings a buyer and seller together. It also requires enhancement to Grid job submission systems.

In order to access the service, the GSC has to present credentials to the GSP. In the context of GridBank we consider such credentials to be a payment instrument that GSC obtains from the GridBank. The payment instrument contains GSP's identification (the Certificate Name) and GridBank account details.

Grid Service Consumer (GSC) initiates the process by negotiating service rates with GSP's Grid Trade Service (GTS). Upon mutual agreement on service rates, Grid Resource Broker (GRB) contacts GridBank Payment Module (GBPM) with request for access to the GSP.

GBPM obtains a payment instruction, for example, a GridCheque, from the GridBank Server. It then forwards the payment and agreed service rates to the GridBank Charging Module (GBCM). GBCM confirms that the payment and service rates offer are valid. If so, the module decides to grant access to GSC. The access to the GSP is accomplished in the usual fashion by using Globus Toolkit. This implies that GSP needs a local account to be associated with the GSC [15]. In order to achieve scalability, only several accounts are maintained by the system and are dynamically allocated to GSCs as they request for service.

GSP maintains a pool of template accounts [16]. These accounts are local system accounts that are not associated with any particular user. When a GSC contacts GSP to execute some application, provided GSC presents a well-formed payment instrument, GSP dynamically assigns one of the template accounts from the pool of free accounts. GSC's Certificate Name is temporarily mapped to the local account (*in grid-mapfile* [15]) to indicate the dynamic relationship between the account and current user. GSP retains the fine-grained access control to its resources by specifying permissions on the template accounts.

When application has finished execution, Grid Resource Meter collects information about resource usage against the local account and forwards it to the GBCM. GBCM calculates total cost based on the resource usage and the mutually agreed service rates. These charges are recorded against the GSC associated with the local account. GBCM then removes the association by deleting the entry corresponding to GSC in the grid-mapfile [15] and returning the local account to the pool of free accounts. GBCM then forwards the payment instrument (e.g. GridCheque) along with the Resource Usage Record to the GridBank for processing.

## 3 Grid Bank System Architecture

### 3.1 Payment strategies

We employed a layered and modular architecture for GridBank to leverage existing technologies and manage them as separate components (Figure 3). This approach allows different payment schemes such as digital cheques, coins, hash chains and other [1,17,18,19,21,22] to be easily integrated with other GridBank modules. Depending on charging strategy, GSC and GSP can select appropriate protocol and exchange service for currency. The charging policies include [2,4] :

1. Pay before use
2. Pay as you go
3. Pay after use

The first policy is appropriate for services that have a fixed cost, for example, to access a directory service. A simple funds transfer protocol is designed to enable GSC to request funds transfer with the confirmation send to GSP. GSC establishes secure connection with GridBank to provide account details of GSC and GSP as well as amount and URL of GSP. GridBank performs the funds transfer and sends the confirmation to the specified URL of the GSP via another secure channel.

The second policy might be used to eliminate unnecessary trust relationships between GSCs and GSPs. A hash chain scheme based on PayWord [21,22] would allow service consumers to dynamically pay service providers for CPU time or per each computation result delivered.

The third payment strategy emulates credit card payment model. When the service charge is unknown beforehand, GSC forwards a payment order in the form of a digital cheque to GSP. The cheque is made out to GSP so no one else can redeem it. After computation has finished, GSP calculates total cost and forwards the cheque along with resource usage record to GridBank for processing. This can be done in batches. Such scheme is based on NetCheque protocol [18,22] and relies on public key cryptography.

Secure connections between parties involved in a transaction are provided by the Security Layer (see Figure 3). In our implementation we chose Public Key Infrastructure (PKI) since it is already provided by Globus Toolkit's GSI (Globus Security Infrastructure) [11,15]. Client authentication and authorization are part of GSI. Secure communication between all participants of any GridBank transaction use Globus I/O API, which implements GSS (Generic Security Service) API. GSS API also provides symmetric data encryption based on SSL technologies to securely exchange sensitive financial information.

### 3.2 Server architecture

GridBank's server architecture (see Figure 3) consists of several modules that can be enhanced or replaced without affecting other modules. These modules are organized into three layers. Accounts Layer deals with database and account operations. Payment Protocol Layer defines payment schemes, message formats and communication protocols. Security Layer ensures that any messages passed to Payment Protocol and consequently invoking operations in Accounts Layers are authentic.

*GB database* module is a relational database that stores account and transaction information.

*GB Accounts* is the core module interacting with the GB database. It provides functions for basic account operations such as creation of accounts, requesting and updating account details, transfer of funds from one account to another, locking funds and transfer from locked funds. This module is independent of payment scheme, protocols used and underlying security model. Its purpose is to perform database operations that deal with manipulating and managing GridBank's database.

*GB Admin* module provides account management such as deposit, withdrawal, change credit limit, cancel transfers and close account functions. These functions are performed by GridBank's administrators who are responsible for transferring real money to and from clients. In the future, this process can be automated by using other payment systems such as PayPal or credit card transactions by importing data from those systems. Administrators have a privileged access and the credentials for such access are checked by GB Administration module.

The payment protocol layer represents payment schemes and associated protocols that interact with GB Accounts. The protocols are described in section 3.1. *GridCheque Protocol* module implements the pay-after-use scheme. *GridHash Protocol* module implements pay-as-you-go payment strategy. Pay-before-use protocol does not involve generation of any payment instruments. Transaction is performed on-line; secure connections guarantee authenticity of participants. Any other payment scheme that defines its own data structures and communication protocol can be added without need to modify GB Accounts or GB Security modules.

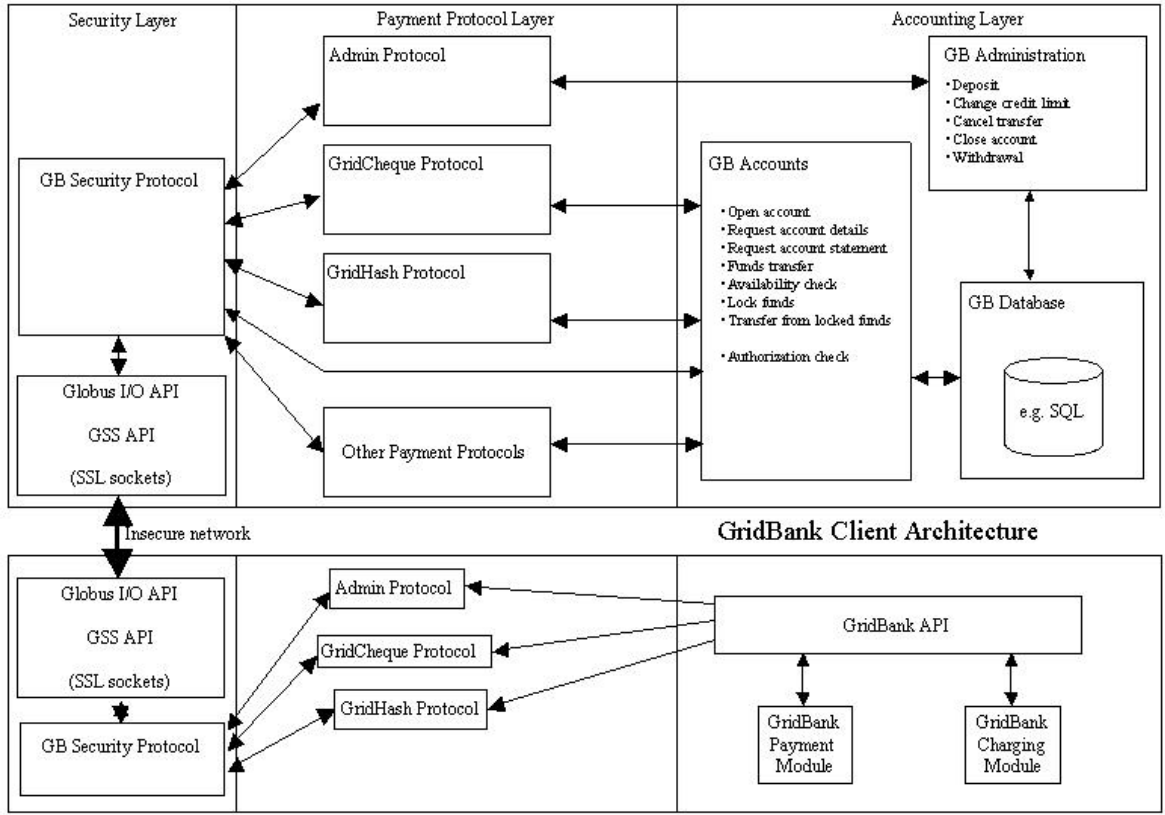

Figure 3: GridBank Server Architecture

*GB Security Protocol* module performs authentication and authorization of GridBank's clients. Authentication process uses Generic Security Services (GSS) API, which is implemented by Globus Toolkit I/O module [14]. It is based on Public Key Infrastructure using X509v3 certificates [6,11]. Certificates can be issued by the Globus Certificate Authority (CA) [15]. Alternatively, GridBank can set up its own CA. Once clients are authenticated, the certificate subject name is retrieved using Globus I/O API and is checked against the database (Figure 3). If the subject name appears either in the accounts or in administrator tables, then the client is authorized to establish a connection. Otherwise connection is refused, and this provides a mechanism to limit denial-of-service attacks. Clients simply cannot send any requests before a connection is established. Only clients with existing account or administrator privilege are authorized and connected.

The GSPs entities, such Grid Trader working with Grid Meter, are responsible for enforcing accounting by billing the GSC account. For example, Nimrod-G broker has an entity called Grid-Agent responsible to executing the user job on Grid node, also keeps track of resource consumption, which can used by the Trade Server to enforce accounting.

### 3.3    Client architecture

The Security Layer is identical to the server. The Protocol Layer has same protocol modules as the server with corresponding client functionality. GridBank API provides an interface to the Protocol layer, which is responsible for obtaining payment instruments or performing direct transfers. GridBank Payment Module and GridBank Charging Module interface to GridBank API module to invoke GridBank operations. The GBPM requests payments whereas GBCM redeems payments.

### 3.4  Payment Guarantee

When a chargeable service has a fixed price as in Pay Before Use strategy or the client obtains hash chains, there are no issues regarding availability of funds; a client could never overspend since his/her account is checked and decremented beforehand. On the other hand, when a credit card approach is taken as in Pay After Use strategy when the total cost is not known beforehand, clients can easily spend more than they have in the account (even taking into account credit limit this is also an issue if cost exceeds available balance together with credit limit). To guarantee payment when issuing GridCheques, GridBank will have to lock a certain amount of funds for the cheque to be valid. The exact amount will depend on the budget constraint set with the GRB. Each GSP will receive a cheque with a reserved amount, which is transferred to the "locked" balance of the GSC's account.

## 4  Grid Bank Operating Models

### 4.1  Co-operative model

In co-operative computing environments, all participants both consume and provide services; when participants provide services, they earn credits. They can use/spend those credits to get access to other resources within the community when needed [2,4]. GridBank supports this form of resource bartering. Each participant may be initially is allocated a certain amount of credits. The amount depends on the value of the resource the participant owns. How the value is determined must be decided by the community and is outside the scope of GridBank.

To achieve price equilibrium, supply and demand need to be carefully regulated in such a way so that GSPs are paid approximately as much currency as they will use to access other Grid services. Otherwise the whole environment will end up in a state where some participants, who do not require any services, have all the money while others who want to access GSPs have none. A community based resource valuation and pricing authority is needed to control prices of Grid resources.

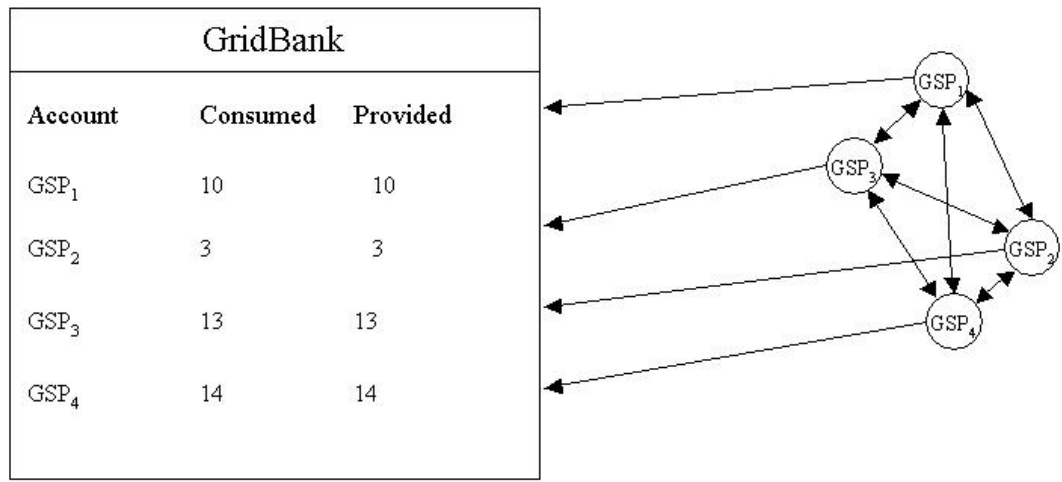

Figure 4: Co-operative Resource Sharing

**Co-operative Resource Sharing Use Case:** Four Grid Services Providers, who are consumers as well, use GridBank to record how much of other resources they have consumed. To maintain price equilibrium the participants should ensure that they provide approximately the same value of resources. In Figure 4, the GridBank accounts show how much of Grid currency each client have consumed and provided. Although computations on some resources are faster because of better hardware, the slower resources have to compensate by running longer.

## 4.2 Competitive model

In competitive computing environments GSPs are allowed to solicit any price [2,4]. However, to attract customers they need estimates of market value of their resources. GridBank's transaction history can assist in deciding how much a computational service is worth. Such transaction history is confidential and cannot be disclosed as is. Therefore GridBank would receive a description of the resource, process the information in its database regarding prices paid for resources of similar type, and then produce an estimate. The simplest approach to compare resources is to consider hardware parameters such as processor speed, number of processors, amount of main memory and secondary storage, network bandwidth, etc. More sophisticated methods are to be considered as GridBank evolves.

## 5 Data Structures and APIs

GridBank's database consists of the following records, defined using MySQL data types.

### 5.1 Database Record Types

ACCOUNT RECORD

- AccountID : VARCHAR(16) (imitates real world account numbers: bank number-branch number-account number. E.g. 01-0001-00000001)
- CertificateName : VARCHAR(150) (X509v3 Certificate Name - globally unique client identifier)
- OrganizationName : VARCHAR(30) (optional)
- AvailableBalance : FLOAT (represents Grid credits/dollars/money units)
- LockedBalance : FLOAT (to guarantee payment for jobs that already have started)
- Currency : VARCHAR(10) (e.g. GridDollar, USD, AUD)
- CreditLimit : FLOAT (default is 0)

TRANSACTION RECORD

- TransactionID : BIGINT(20) UNSIGNED (Unique transaction identifier)
- Type : VARCHAR(10) (Possible types are: Deposit, Withdrawal, Transfer. In case of transfer, there is a corresponding transfer record with the same TransactionID)
- Date : TIMESTAMP(14) (Date when transaction was committed to database)
- Amount : FLOAT (Amount of transaction: if withdrawal or transfer from the account, then the amount is negative)

TRANSFER RECORD

- TransactionID : BIGINT(20) UNSIGNED (Same as in the corresponding transaction record)
- Date : TIMESTAMP(14) (Date when the transfer was committed to database)
- DrawerAccountID : VARCHAR(16) (GSC Account ID)
- Amount : FLOAT (Transfer amount: always positive)
- RecipientAccountID : VARCHAR(16) (GSP Account ID)
- ResourceUsageRecord : BLOB (contains RUR in a binary format)

**NOTE:** The Resource Usage Record format needs to be defined. However, whatever format is chosen (e.g. XML), GridBank stores RUR in binary format. This approach will achieve greater flexibility as the RUR can be independently defined by the Grid sites and they do not have to conform to any standard, although they should. Grid Resource Meter Module can then perform translations from one record format into another.

The format of RUR is being defined by the Grid community effort [13,20]. Currently, the following items are being associated with RUR:

- User details
  - Host name / IP address
  - Certificate Name (Grid-wide unique ID of GSC)
- Job details

- Job ID (Nimrod-G? Or local process id on the resource? Or GRID global unique id?)
- Application name
- Job start date (includes time)
- Job end date
* Resource details
  - Host name / IP address
  - Certificate Name (Grid-wide unique ID of GSP)
  - Host type (e.g. Cray; optional)
  - Local job id (local OS process id to settle disputes about resource consumption)
  - Wall clock time + price per time unit (e.g. Per second)
  - CPU time + price per time unit
  - Main memory + price per time unit
  - Secondary storage + price per time unit
  - Network activity + price per time unit
  - Software service + price per time unit
* Total price per time unit
* Job Cost ( = (end date - start date) * total price per time unit )

## 5.2 GridBank API

GridBank API module (Figure 3) defines the following operations for GridBank Payment Module and GridBank Charging Module.

- Create New Account

    Input: Client's Certificate (Certificate is checked for authenticity; if legitimate, then Certificate Name is extracted)

    Output: AccountID

- Request Account Details / Check Balance

    Input: AccountID

    Output: ACCOUNT RECORD

- Update Account Details

    Input: ACCOUNT RECORD (First, Request Account Details operation returns current record, a client amends it and invokes this operation to update the record. Only CertificateName and OrganizationName can be modified.)

    Output: Confirmation

- Request Account Statement

    Input: Account ID, Start Date, End Date

    Output: ACCOUNT RECORD, TRANSACTION and TRANSFER RECORDS (from Start to End Dates)

- Perform Funds Availability Check (For protocols where payee requires confirmation of a certain amount; the amount is transferred into locked balance for guarantee)

    Input: Account ID, Amount
    Output: Confirmation

- Request Direct Transfer

            Input: From AccountID, To AccountID, Amount, RecipientAddress

            Output: Confirmation sent to RecipientAddress

- Request GridCheque

            Input: AccountID, Amount

            Output: GridCheque

- Redeem GridCheque

            Input: GridCheque, Resource Usage Record

            Output: Confirmation

- Request GridHash chain

            Input: AccountID, Amount

            Output: GridHash chain

- Redeem GridHash chain

            Input: GridHash chain, Resource Usage Record

            Output: Confirmation

### 5.2.1 GridBank Admin API

- Deposit funds (GridBank administrator receives funds via existing credit/debit/smart card payment systems, and deposits same amount into GridBank account)

            Input: AccountID, Amount

            Output: Confirmation

- Change credit limit

            Input: Credit limit amount

            Output: Confirmation

- Cancel Transfer

            Input: Drawer AccountID

            Output: Confirmation

- Withdraw (The withdrawn funds will be transferred to an actual bank account)

            Input: Drawer AccountID

            Output: Confirmation

- Close account and get outstanding balance transferred to another GridBank account or actual bank account.

### 5.3 GridBank Payment Module API

GBPM defines the following API for Grid Resource Broker or as commands for job submission to resources requiring payment.

- Grid Bank Job Submit; similar to Globus Toolkit's "globus-job-submit" [15], but for GridBank-enabled Grid services.

The following are identical to GB API, but are independent of input parameters, which are stored by GBPM.

- Create New Account
- Request Account Details / Check Balance
- Update Account Details
- Request Account Statement

GBCM is an internal module and does not need an API. GSPs can access their account information via GBPM, since both modules are part of GridBank client software on each resource.

## 6      Conclusion and future work

We presented new Grid components in the context of the Grid-wide banking service that will enable participants to engage in global computational economy. Grid Resource Meter extracts resource usage information from the operating system and converts it into a Grid-wide standard form. GridBank Charging Module is responsible for determining legitimacy of payment instruments passed to it by the GridBank Payment Module, setting up and removing (after execution of user application) temporary local accounts, calculating total charge using the Resource Usage Record and the service rates passed by the Grid Trade Service, and redeeming the payment with the GridBank server. GridBank Payment Module receives requests for job execution from the Grid Resource Broker, obtains a payment instrument from the GridBank, forwards the payment to GBCM and submits the job when GBCM notifies GBPM that a local account has been set up. Grid Trade Server negotiates service cost/rates with GRB and provides interface for GBCM to obtain the information. Negotiation protocols are already defined in [2,4].

In the future, GridBank system will be expanded to provide multiple servers/branches across the Grid to achieve scalability in similar manner as the currency servers in NetCash [17] and NetCheque [18] systems. It is precisely for this purpose that GridBank accounts have branch numbers. Each Virtual Organization (VO) [10,12], which is a collaboration of resources, associates a GridBank server that all participants of the organization use. If a GSC is from one VO and GSP is from another, then their respective servers will need to define protocols for settling accounts between the branches. Moreover, if another payment system is introduced to the Grid, then that system can use different bank number and additional protocols can be defined to settle accounts between multiple banks.